\documentclass[aps,preprint,prc, showpacs, showkeys]{revtex4-1}

\usepackage{dcolumn}% Align table columns on decimal point
\usepackage{epsfig,amssymb,multirow,amsmath, bm}
\usepackage{xcolor,soul}
\usepackage{hyperref}
\newcommand{\sgn}{\mathop{\mathrm{sgn}}}
\begin{document}

%\preprint{}

\title{Saturated symmetric nuclear matter in strong magnetic fields} %Title of paper



\author{J.P.W. Diener}
	\email{jpwd@sun.ac.za}
 \affiliation{National Institute for Theoretical Physics (NITheP), Stellenbosch 7600, South Africa}
 \affiliation{Institute of Theoretical Physics, University of Stellenbosch, Stellenbosch 7600, South Africa}
 %\altaffiliation[Also at]{Institute of Theoretical Physics, University of Stellenbosch, Stellenbosch 7600, South Africa}
\author{F.G. Scholtz}%
 \email{fgs@sun.ac.za}
 \affiliation{National Institute for Theoretical Physics (NITheP), Stellenbosch 7600, South Africa}
 \affiliation{Institute of Theoretical Physics, University of Stellenbosch, Stellenbosch 7600, South Africa}

%	\altaffiliation[Also at ]{Institute of Theoretical Physics, University of Stellenbosch, Stellenbosch 7600, South Africa}

\date{\today}

\begin{abstract}
Strongly magnetized symmetric nuclear matter is investigated within the context of effective baryon-meson exchange models. The magnetic field is coupled to the charge as well as the dipole moment of the baryons by including the appropriate terms in the Lagrangian density. The saturation density of magnetized, symmetric nuclear matter $\rho_0(B)$ was calculated for magnetic fields of the order of $10^{17}$ gauss. For the calculated range of $\rho_0(B)$ the binding energy, symmetry energy coefficient $a_4$ and compressibility $K$ of nuclear matter were also calculated. It is found that with an increasing magnetic field $\rho_0(B)$ increases, while the system becomes less bound. Furthermore, the depopulation of proton Landau levels leaves a distinct fluctuating %oscillatory 
imprint on $K$ and $a_4$.
The calculations were also performed for increased values of the baryon magnetic dipole moment. By increasing the dipole moment strength
$\rho_0(B)$ is found to decrease, but the system becomes more tightly bound while the fluctuations %oscillatory behavior of 
in $K$ and $a_4$ persist.
\end{abstract}

 \pacs{26.60.Kp, 13.75.Gx, 21.65.Cd, 21.65.Ef}
%\pacs{26.60.Kp, 21.30.Fe, 21.60.Jz, 21.65.Cd}% insert suggested PACS numbers in braces on next line
\keywords{magnetized nuclear matter, relativistic mean-field, symmetry energy, compressibility}
\maketitle %\maketitle must follow title, authors, abstract and \pacs
\section{Introduction}
Extremely strong magnetic fields cannot be produced in laboratories, but these conditions can be found in stellar environments. All stars are magnetized and the strongest magnetic fields are found in the stars generically known as neutron stars. 
They are observed as rapidly rotating, strongly radio-emitting objects called pulsars with magnetic fields of between $10^8$ and $9\times10^{13}$ G \cite{pnas}. However, x-ray or $\gamma$-ray emitting pulsars are also observed. They are assumed to be highly magnetized neutron stars with magnetic fields between 10$^{14}$ and 10$^{15}$ G and are called magnetars. For a review of their properties see Ref.\ \cite{chap14}. \\
\\
The current model of magnetars assumes that the star's magnetic field is formed in the interior of the progenitor through dynamo action \cite{T+D93}. Since the observed magnetic field strengths are those on the surface of the star, the interior magnetic field may well be larger. Using realistic equations of states as well as a general relativistic description of rotating magnetized stars Kuichi and Kotake \cite{KandK} calculated that a magnetar surface magnetic field of about $10^{16}$ G would increase to a maximum of over $10^{17}$ G in the interior. Frieben and Rezzolla, using a similar approach, found the average magnetic field in the magnetar interior to be of the order of $10^{17}$ G while the maximum value would be between $3.26\times 10^{17}$ G and $8.05\times 10^{17}$ G, depending on the equation of state \cite{FandR}. The current consensus seem to be that a magnetar cannot sustain a magnetic field larger than between $10^{18}$ and $10^{19}$ G, as summarized in Ref.\ \cite{NS1}.\\
\\
It is an open question whether, and in what way, these strong magnetic fields can influence the properties of the matter in the magnetar interior. One assumption that is made about the neutron star interior is that it contains nuclear matter in charge and beta equilibrium \cite{csg}. Pe\~na Arteaga {\em et al.} concluded that, depending on the specific nucleus, magnetic field strengths of about $5\times 10^{16}$ G and larger could affect the nuclear shell structure \cite{penaA}. Although nuclear matter is homogeneous and thus does not have a shell structure, the expected range of the magnetic field in the magnetar interior could influence its various other properties.\\
\\
Due to the nuclear interaction being short ranged it saturates at higher densities and favors isospin symmetry \cite{csg}. However, owing to the condition of charge neutrality imposed on the neutron star interior and the short half-life of free neutrons, perfectly symmetric matter would not occur in the neutron star interior. On the other hand, the equation of state and characteristics of high density asymmetric nuclear matter is unknown (see Ref.\ \cite{a4rev} for a recent review). Thus, in order to get a first approximation of the behavior of nuclear matter in strong magnetic fields, we turn our attention to symmetric nuclear matter, since it has definite properties which are related to finite nuclei and nuclear matter \cite{csg}. \\
\\
In this paper we investigate the impact that a very strong, external magnetic field has on the properties of cold, saturated symmetric nuclear matter for a range of the baryon magnetic dipole moment strengths. In particular we calculate the saturation density as a function of the magnetic field strength. Furthermore %Other properties that will be investigated are 
the binding energy, the compressibility, and the symmetry energy, all at saturation, will also be investigated. These characteristic properties have been established as indicators of the behavior of dense nuclear matter and are generally used to constrain nuclear matter models, such as quantum hadrodynamics (or QHD, which is also known as the Walecka-model) \cite{walecka} and its extensions \cite{serot}.\\ 
\\
In QHD the nuclear interaction between the baryons stems from various meson exchanges, where each exchange describes a different feature of the nuclear interaction. The free parameters are the meson-baryon coupling constants, as well as the meson self-coupling constants. We will make use of the QHD1 \cite{walecka}, NL3 \cite{NL3}, FSU (or FSUGold) \cite{FSU1}, and the more recent IU-FSU \cite{IUF} parameter sets. QHD1 is one of the earliest parameter sets and includes only the scalar sigma and vector omega mesons in its description \cite{walecka}. NL3 also includes the isovector rho meson and introduces a self-coupling in the scalar meson field to improve the description of nuclear matter \cite{NL3}. FSU was parametrized to investigate, among others, the nuclei away from nuclear symmetry. It introduces a self-coupling in the vector field, as well as a coupling between the vector and isovector mesons \cite{FSU1}. FSU and IU-FSU share the same couplings, but the latter was constrained to also satisfy astrophysical requirements \cite{IUF}. Despite the fact that all QHD parameter sets are constrained to reproduce the same nuclear matter properties at nuclear saturation, they have very different behavior at densities above saturation \cite{serot}. \\
\\
In order to perform calculations the system is approximated using the relativistic mean-field (RMF) approximation. In the RMF approximation the meson fields operators are replaced by their ground state expectation values and become classical fields \cite{walecka}. As noted in Ref.\ \cite{NS1}, the RMF approximation is at best a phenomenological description of nuclear matter. The RMF approximation is very good when the meson interaction length is much larger than the spacing between the baryons. However, for the densities at which the approximation is applied the distance between baryons is actually of the order of the meson interaction length. Despite this inconsistency, the calculated RMF nuclear properties have shown good agreement with experimentally known properties of nuclei and nuclear matter \cite{serot}.\\
\\
Various aspects of magnetized nuclear matter have already been investigated using QHD in the RMF approximation, most recently 
by Dong {\em et al.} \cite{dong}. In the latter work the density-dependence of the symmetry energy of magnetized matter was investigated with the FSU parameter set at various densities as well as proton and neutron ratios, while also adjusting some of the coupling strengths in the parameter set. The authors concluded that the parabolic isospin dependence on the energy per nucleon remains valid for strong magnetic fields. An overview of previous studies is also provided in Ref.\ \cite{dong} and references therein.\\
\\
Casali {\em et al.}\ \cite{poa4} investigated the impact of magnetic fields of $10^{17}$ and $10^{18}$ G on the symmetry energy coefficient at densities below nuclear saturation using the NL3 and FSU QHD parameter sets. They found that the Landau levels give rise to discontinuities in the symmetry energy and influence the composition of the neutron star crust. They also investigated the effect of including the coupling between the magnetic field and the baryon dipole moments and concluded that it will only be appreciable in very strong magnetic fields at subsaturation nuclear densities. However, it would appear that the contribution of the magnetic field was not included in the energy density of the neutron star matter. As pointed out by Broderick {\em et al.} in Ref.\ \cite{Brod00}, the magnetic contribution should be included since it influences the equation of state and thus the composition of the matter. \\
\\
Broderick {\em et al.} \cite{Brod00} were also the first to point out the importance and impact of including the coupling between the dipole moment of the baryons and the magnetic field, in addition to the coupling of the proton's charge to the magnetic field, in the description of magnetized matter. In Ref.\ \cite{Brod00} this coupling is referred to as the {\em anomalous magnetic moment} or ``AMM'' coupling with the coupling strength of the applicable baryon's magnetic dipole moment. We believe this to be a somewhat misleading term since baryons are not point particles like electrons, but have an internal structure of quarks and gluons. The anomalous contribution to the electron dipole moment arises from higher order photon couplings to the electron charge. On the other hand, the baryon's dipole moment (partially or fully, depending on whether the baryon is charged or not) arises from the photons coupling to the baryon's charged internal structure. Thus these ``anomalous'' contributions to the baryon dipole moment are not higher order contributions of the electromagnetic coupling, but an expression of the fact that the baryons have internal structure. However, not to confuse the reader we will also adopt this naming convention.\\
\\
Due to this particular origin of the baryon magnetic dipole moment, more than anecdotal evidence would suggest that it should not be constant under all conditions. One would expect that the internal baryon structure would be affected by the baryon density, especially at high densities. Since this structure is the origin of the dipole moment, by extension the baryon magnetic dipole moment would also be influenced. As discussed by Berryman \cite{berryman}, experimental investigations would also suggest the proton dipole moment is altered at higher densities. It is shown in Ref.\ \cite{berryman} (based on data from Ref.\ \cite{stone}) that the dipole moment of copper, which has one proton outside the closed Z=28 proton shell, with an even number of valence neutrons increases by about 50$\%$ over a mass number range of 10. Furthermore, Ryu {\em et al.}\ in Ref.\ \cite{Ryu} investigated the neutron star equation of state with density-dependent dipole moments for the baryon octet using the quark-meson coupling (QMC) models and extensions thereof. They found that the neutron star equation of state is dependent on both the strength of the magnetic field as well as that of the baryon magnetic dipole moment. Unfortunately the density-dependence of the baryon dipole moment at high densities is not known, either experimentally or theoretically, which complicates the calculations.\\
\\
To achieve our stated goals the following calculations were performed. First, the saturation density of symmetric nuclear matter $\rho_0(B)$ was calculated at the density that minimizes the binding energy per nucleon for a range of magnetic field strengths $B$. Then the compression modulus and the symmetry energy coefficient were calculated at these densities. Since the density dependence of the baryon magnetic dipole moment is not known, these calculations were repeated for a range of values of the baryon magnetic dipole moment. Our results are presented and discussed at the end of the paper. To start off we present an overview of our formalism.
\section{Formalism}
The interacting part of the QHD RMF Lagrangian for magnetized matter together with the free electromagnetic component is \cite{dong, IUF}
%\begin{widetext}
\begin{eqnarray}
%\begin{align}
\label{LBB}
%		\begin{split}
				{\cal L}_{int}&=&%\nonumber\\
				\bar{\psi}
				\left[g_{s}\phi_0-\gamma^{\mu}\left(q_b\frac{1+\tau_3}{2} A_\mu + g_{v}V_0 
				+ \frac{g_\rho}{2}{\tau_3}{b_0}\right)
				\right]\psi
				%- \frac{1}{2}m_s^{2}\phi^{2} 
				- \frac{\kappa}{3!}\big(g_s\phi_0\big)^3 - \frac{\lambda_\phi}{4!}\big(g_s\phi_0\big)^4 \nonumber \\
				&&
				%+ \frac{1}{2}m_\omega^{2}{V_0}^2 + 
				+\frac{\zeta}{4!}\big(g_v{V_0}\big)^4 %\\
				%&&%+\frac{1}{2}m_{\rho}^{2}b_0^2
				+\, \Lambda_v\big(g_v{V_0}\big)^2
				\left(g_\rho b_0\right)^2%\nonumber \\
				%&&
				-\bar{\psi}\frac{g_b}{2}F^{\mu\nu}\sigma_{\mu\nu}\psi-\frac{1}{4}F^{\mu\nu}F_{\mu\nu},%\nonumber,
%			\end{split}
%\end{align}	
\end{eqnarray}
%
%\begin{eqnarray}
%%\begin{align}
%\label{LBB}
%%		\begin{split}
				%{\cal L}_{int}&=&%\nonumber\\
				%\bar{\psi}
				%\left[g_{s}\phi_0-\gamma^{\mu}\left(q_b\frac{1+\tau_3}{2} A_\mu + g_{v}V_0 
				%+ \frac{g_\rho}{2}{\tau_3}{b_0}\right)
				%\right]\psi\nonumber \\
				%&&
				%%- \frac{1}{2}m_s^{2}\phi^{2} 
				%- \frac{\kappa}{3!}\big(g_s\phi_0\big)^3 - \frac{\lambda_\phi}{4!}\big(g_s\phi_0\big)^4 %\nonumber \\
				%%&&
				%%+ \frac{1}{2}m_\omega^{2}{V_0}^2 + 
				%+\frac{\zeta}{4!}\big(g_v{V_0}\big)^4 \\
				%&&%+\frac{1}{2}m_{\rho}^{2}b_0^2
				%+\, \Lambda_v\big(g_v{V_0}\big)^2
				%\left(g_\rho b_0\right)^2%\nonumber \\
				%%&&
				%-\bar{\psi}\frac{g_b}{2}F^{\mu\nu}\sigma_{\mu\nu}\psi-\frac{1}{4}F^{\mu\nu}F_{\mu\nu}\nonumber,
%%			\end{split}
%%\end{align}	
%\end{eqnarray}
%\end{widetext}
where 
$\psi=\left[
			\begin{array}{c}
				 \psi_p\\
				\psi_n
			\end{array}	
		\right]$ 
is the isodoublet baryon field operator where subscript $p$ and $n$ indicate the proton and neutron components, while $\phi_0$, $V_0$, and $b_0$ indicate the scalar, vector, and isovector mesons. The mesons couple to the baryons via $g_s,\ g_v$, and $g_\rho$ while $\kappa,\ \lambda_\phi,\ \zeta$, and $\Lambda_v$ are the meson self-coupling strengths. The values of the couplings and meson masses are given in Table \ref{tab:coup}. 
	\begin{table*}
	\centering
		\begin{tabular}{lcccccccccc}
			\hline\hline
			Model &$m_s$& $m_\omega$& $m_\rho$ & $g_s^2$ & $g_v^2$ & $g_\rho^2$& $\kappa$ & $\lambda_\phi$ & $\zeta$ & $\Lambda_v$ \\
			\hline
			QHD1 & 520 & 783 & 0.0 & 109.6 & 190.4 & 0.0 & 0.0 & 0.0 & 0.0 & 0.0 \\
			NL3 & 508.194 & 782.501  & 763.000 & 104.3871 & 165.5854 & 79.6000 & 3.8599 & -0.015 905 & 0.00 & 0.00 \\
			FSU & 491.500 & 782.500  & 763.000 & 112.1996 & 204.5469 & 138.4701& 1.4203 & +0.023 762  & 0.06 & 0.030 \\
			IU-FSU & 491.500 & 782.500  & 763.000 & 99.4266 & 169.8349 & 184.6877 & 3.3808 & +0.000 296  & 0.03 & 0.046 \\
			\hline
		\end{tabular}
		\caption{\bf Coupling constants of different QHD parameter sets from Refs.\ \cite{serot} and \cite{IUF}.  All coupling constants are dimensionless, except for $\kappa$ which is given in MeV. The baryon mass $m$ is taken as 939 MeV, while $m_s$, $m_\omega$ and $m_\rho$ are given in MeV.}
	\label{tab:coup}
\end{table*}
Furthermore, $\tau_3$ is the isospin operator, 
		$
			A^\mu=(0,0,Bx,0)%\label{Amu},
		$
		with $B=|\bm B|$, while $\sigma^{\mu\nu} = \frac{i}{2}\left[\gamma^\mu,\gamma^\nu\right]$ are the generators of the Lorentz group \cite{gross}. The strength of the coupling between the baryons and $A^\mu$ is the baryon charge $q_b=\left[
			\begin{array}{cc}
				 q_p&0\\
				0&q_n
			\end{array}	
		\right]$, while $g_b=\left[
			\begin{array}{cc}
				 g_p&0\\
				0&g_n
			\end{array}	
		\right]$ is the strength of the coupling between the magnetic and baryon fields (with units of the baryon magnetic dipole moment). Under normal conditions the proton dipole moment is $2.793\,\mu_N$ while the neutron's is $-1.913\,\mu_N$ (expressed in units of the nuclear magneton $\mu_N$) \cite{PDGmuon}. In Ref.\ \cite{dienerPhD} it is shown that the values of $g_n$ and $g_p$ equal to
	\begin{eqnarray}\label{gb0}
		g_n=\frac{1.913}{2}\mu_N=g_n^{(0)}%\label{gn0}\\
		\ \mbox{ and }\ 
		g_p=-\frac{0.793}{2}\mu_N= g_p^{(0)}%\label{gp0}
	\end{eqnarray}
reproduce the normal values of the dipole moments in the nonrelativistic limit.
To adjust the baryon dipole moments by a factor of $x$, $g_n$ and $g_p$ should become \cite{dienerPhD}
\begin{subequations}
	\begin{eqnarray}\label{gb}
		g_n&=&\frac{1.913x}{2}\mu_N=x g_n^{(0)},\mbox{ and }\label{gbp}\\
		g_p&=&-\frac{2.793x-2}{2}\mu_N = xg_p^{(0)}.\label{gbn}
	\end{eqnarray}
\end{subequations}
The second and ninth terms of ${\cal L}_{int}$ (\ref{LBB}) can be expanded to 
\begin{eqnarray}
	-q_p\bar{\psi}_p\gamma^\mu A_\mu\psi_p-\frac{g_p}{2}\bar{\psi}_p{\bm\Sigma}\cdot{\bm B}\psi_p
	-\frac{g_n}{2}\bar{\psi}_n{\bm\Sigma}\cdot{\bm B}\psi_n\label{coupl2},
\end{eqnarray}
where 
$	 {\bm \Sigma}= 
		\left[
			\begin{array}{cc}
				 {\bm \sigma} &0\\
				0 &  {\bm \sigma}
			\end{array}	
		\right]
$ are the baryon spin operators. 
The first term in (\ref{coupl2}) leads to the well-known Landau problem, where the single particle proton energy spectrum is quantized and consecutive levels differ in energy by factor of $|q_p B|$. The last two terms represent the ``AMM'' coupling between the magnetic field and the dipole density $\bar{\psi}{\bm\Sigma}\psi$ of each of the baryons. As shown in Ref.\ \cite{dienerPhD} and references therein, the magnetized proton spectrum is 
\begin{eqnarray}\label{Fp}
	%e(k_z,\lambda,n)&=&\sqrt{k_z^2+\left(\sqrt{{m^*}^2+2|q_p B| n}+\lambda  g_p B\right)^2}\nonumber\\&\ &+ g_vV^0+\frac{g_\rho b_0}{2},
	e(k_z,\lambda,n)&=&\sqrt{k_z^2+\left(\sqrt{{m^*}^2+2|q_p B| n}+\lambda  g_p B\right)^2}+ g_vV^0+\frac{g_\rho b_0}{2},
\end{eqnarray}
while the magnetized neutron spectrum is
\begin{eqnarray}\label{Fn}
	%e(\bm{k},\lambda)&=&\sqrt{k_z^2+\left(\sqrt{k_\perp^2+{m^*}^2}+\lambda  g_n B\right)^2}\nonumber\\&\ &+ g_vV^0-\frac{g_\rho b_0}{2},
	e(\bm{k},\lambda)&=&\sqrt{k_z^2+\left(\sqrt{k_\perp^2+{m^*}^2}+\lambda  g_n B\right)^2}+ g_vV^0-\frac{g_\rho b_0}{2},
\end{eqnarray}
where $k_{\bot}^2 = k_{x}^2+k_{y}^2$, $m^* = (m-g_s\phi_0)$ is the reduced mass, $\lambda=\pm 1$, and $n=\left(n'+\frac{1}{2}-\alpha\,\frac{\lambda }{2}\right)$ with $n'=0,1,2,3...$ and $\alpha = \sgn(q_p B)$ are integers labeling the Landau levels.\\
\\
When the magnetic field is orientated in the $z$ direction (in the 3-dimensional Landau problem) only the $k_z$ momentum contributes to the proton energy. As in the nonrelativistic case, the proton spectrum is degenerate and independent of the choice of $A^\mu$ \cite{yobi}. For neutrons the usual spherical symmetry of the ground state is broken by the magnetic field and replaced with cylindrical symmetry.\\
\\
In (\ref{Fp}) and (\ref{Fn}) $\lambda$ indicates the possible orientations of the baryon dipole moment along $\hat{z}$. Since the Pauli matrices do not commute, the Hamiltonian of magnetized matter does not commute with the baryon spin operator $\bm \Sigma$. Hence spin is not a good quantum number and we rather refer to the orientation of the dipole moment to distinguish between the two types of baryons which are usually referred to as spin up or down.  Therefore, the ``AMM'' coupling induces a relative shift in the energy spectrum of both protons and neutrons with opposite values of $\lambda$. \\
\\
For a system consisting of symmetric nuclear matter the density of protons and neutrons must be equal. Furthermore it is assumed to be at zero temperature and so the ground state will consist of completely filled energy levels with energies up to the Fermi energy. The deformation of the neutron Fermi surface can be accounted for in the calculation of any neutron density, or ground state expectation value, by considering the directional momentum dependence of the Fermi energy (see Ref.\ \cite{dienerPhD} for more details). \\
\\
Due to the Landau quantization, any magnetized proton density is expressed as a sum over the occupied Landau levels. The degeneracy of each proton level is incorporated by adding a pre-factor $\frac{|q_p B|}{4\pi^2}$ to the contribution of the level. The prefactor follows from a comparison of the fundamental magnetic flux per particle to the total magnetic flux through the level \cite{yobi} (also see Ref.\ \cite{dienerPhD} and references therein). \\
\\
Consequently the energy density of magnetized symmetric nuclear matter, including the contribution of the magnetic field, is \cite{dienerPhD}
\begin{widetext}
\begin{align}
	\begin{split}\label{epsLL}
		\epsilon 
		\,=\, &\sum_{\lambda,n}\frac{|q_p B|}{4\pi^2}\int e_p(k_z,\lambda,n)\,\Theta\big[\,\mu_p-e_p(k_z,\lambda,n)\big]dk_z
		+ \sum_\lambda\int\frac{d{\bm k}}{(2\pi)^3}\,e_n(\bm{k},\lambda)\,\Theta[\,\mu_n-e_n(\bm{k},\lambda)] \\
		&
		+ \frac{1}{2}m_s^{2}\phi^{2}_0 +\frac{\kappa}{3!}\big(g_s\phi_0\big)^3
		+ \frac{\lambda_\phi}{4!}\big(g_s\phi_0\big)^4
		 - \frac{1}{2}m_\omega^{2}V_0^2 - \frac{\zeta}{4!}\big(g_vV_0\big)^4 - \frac{1}{2}m_{\rho}^{2}b_0^2 - 
		\Lambda_v\big(g_vV_0\big)^2\big(g_\rho b_0\big)^2+\frac{1}{2}B^2,
	\end{split}
\end{align}
\end{widetext}
where $\Theta$ is the Heaviside step function. A simplified expression for $\epsilon$ can be found in Ref.\ \cite{Brod00}.
\section{Results}
\begin{figure*}
	\makebox[\textwidth][c]{%\centering	
  \begin{tabular}{ll}
    \includegraphics[width=.5\textwidth]{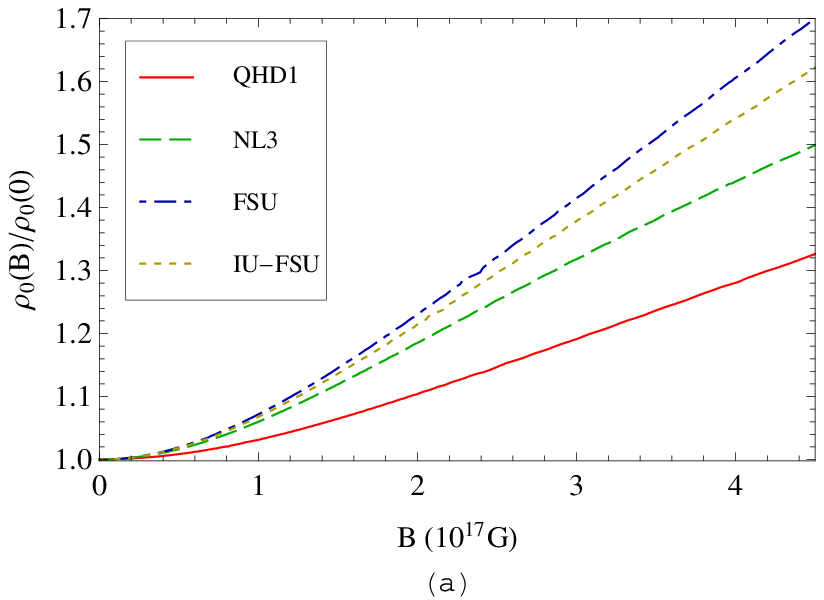}&
    \includegraphics[width=.5\textwidth]{{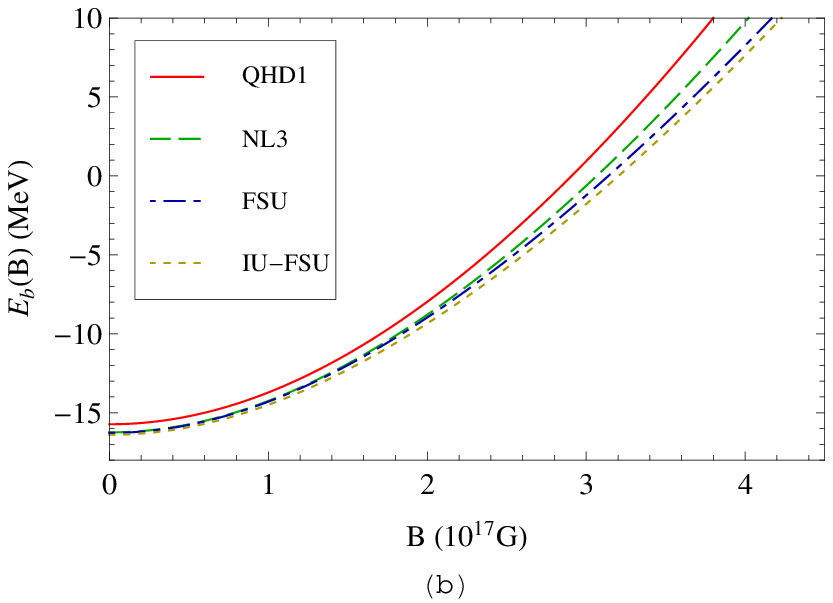}}
  \end{tabular}}
	\caption{ (Color online) Plots of (a) the saturation density normalized with regards to the $B=0$ values, and (b) the binding energy at saturation for different QHD parameter sets with $g_b=g_b^{(0)}$.}\label{fig:BE(B)}	
\end{figure*}

\begin{figure*}
	\makebox[\textwidth][c]{%\centering	
  \begin{tabular}{ll}
    \includegraphics[width=.5\textwidth]{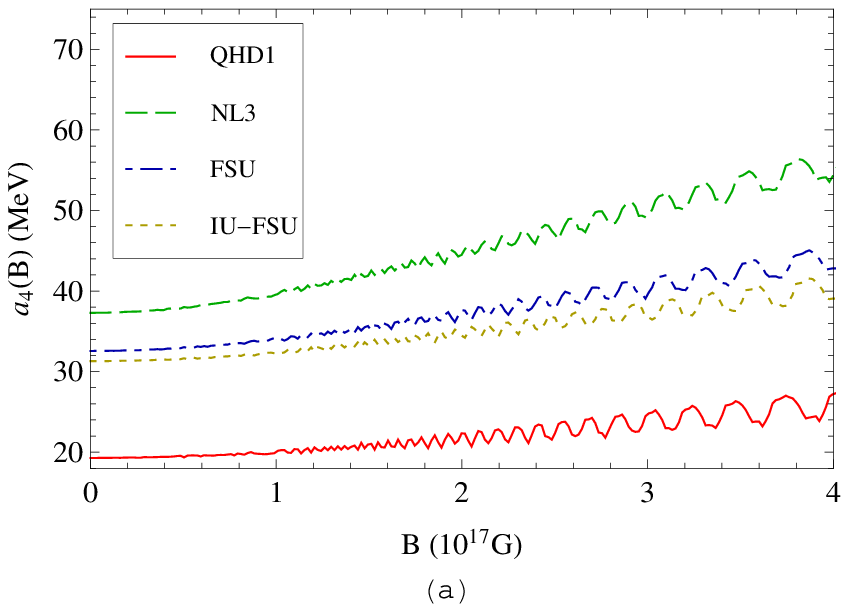}&
    \includegraphics[width=.5\textwidth]{{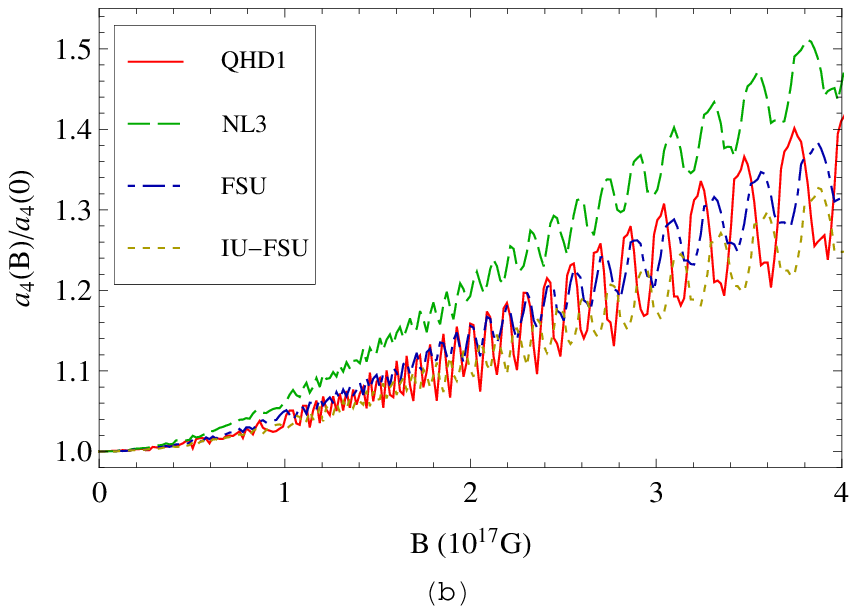}}
  \end{tabular}}
	\caption{ (Color online) The nonnormalized (a) and normalized (b) values of $a_4$ of magnetized nuclear matter for different QHD parameter sets.}
	\label{fig:a4(B)}
	\end{figure*}

\begin{figure*}
	\makebox[\textwidth][c]{%\centering	
  \begin{tabular}{ll}
    \includegraphics[width=.5\textwidth]{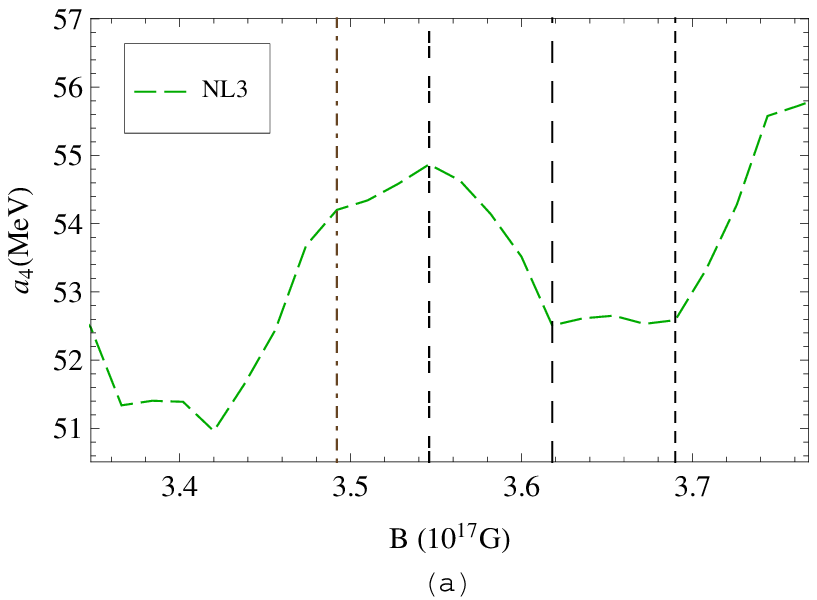}&
    \includegraphics[width=.5\textwidth]{{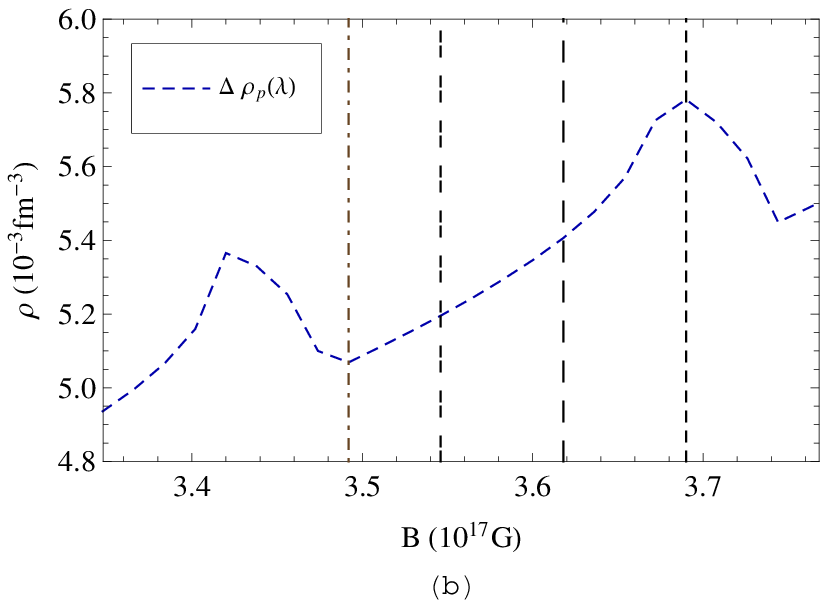}}
  \end{tabular}}
	\caption{ (Color online) Comparison of the NL3 $a_4$ graph (a) and the difference in the proton densities (b). The vertical lines in both plots denote the same values of the magnetic field.}
	\label{fig:a4(B)a}
	\end{figure*}

Since the nuclear interaction is short ranged it saturates at high densities. The saturation density is defined as the density at which the binding energy per nucleon is at a minimum (we follow the convention of the binding energy being negative) and the system is most strongly bound. The binding energy per nucleon is \cite{walecka}
\begin{eqnarray}
	E_b=\frac{\epsilon}{\rho}-m,\label{Eb}
\end{eqnarray}
and most QHD parameter sets reproduce the unmagnetized saturation density $\rho_0(0)$ at $0.148$ fm$^{-3}$ with a binding energy of about $-16.3$ MeV \cite{IUF}. We calculated the saturation density as function of a strong magnetic field with $g_b=g_b^{(0)}$. The results are shown in Fig.\ \ref{fig:BE(B)}.
\\
\\
From these plots we note that $E_b(B)$ behaves very similarly for all parameter sets: As the magnetic field increases the system becomes less strongly bound. At $B\approx3\times 10^{17}$ G, which corresponds to a density of between $1.2$ and $1.4$ times $\rho_0(0)$ (depending on the parameter set), the system becomes unbound. Thus, albeit more weakly bound, the system can accommodate much denser matter as it becomes magnetized. The increase in $\rho_0(B)$ with $B$ was first noted by Chakrabarty {\em et al.} in Ref.\ \cite{chakPRL}. However, in this paper the effect of the ``AMM'' coupling was not studied and the authors found that the system becomes more tightly bound with increasing $B$.
\begin{figure*}[ttb]
	\makebox[\textwidth][c]{%\centering	
  \begin{tabular}{ll}
    \includegraphics[width=.525\textwidth]{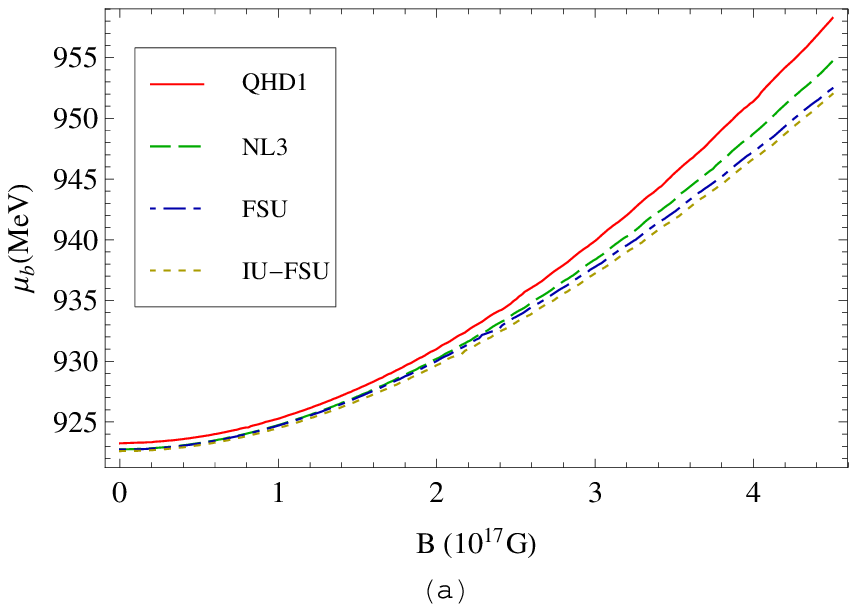}&
    \includegraphics[width=.5\textwidth]{{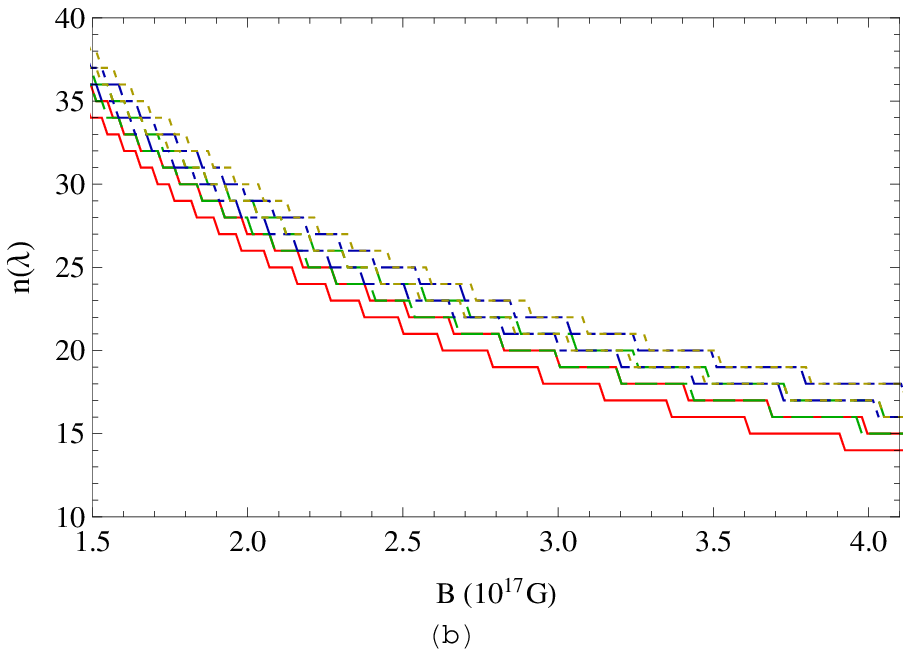}}\\
    \includegraphics[width=.5\textwidth]{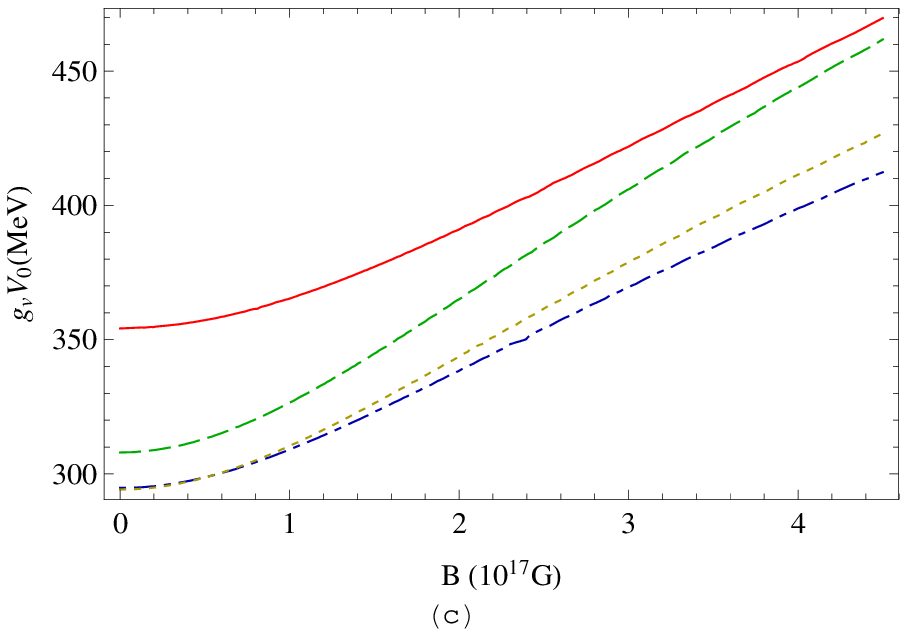}&
    \includegraphics[width=.5\textwidth]{{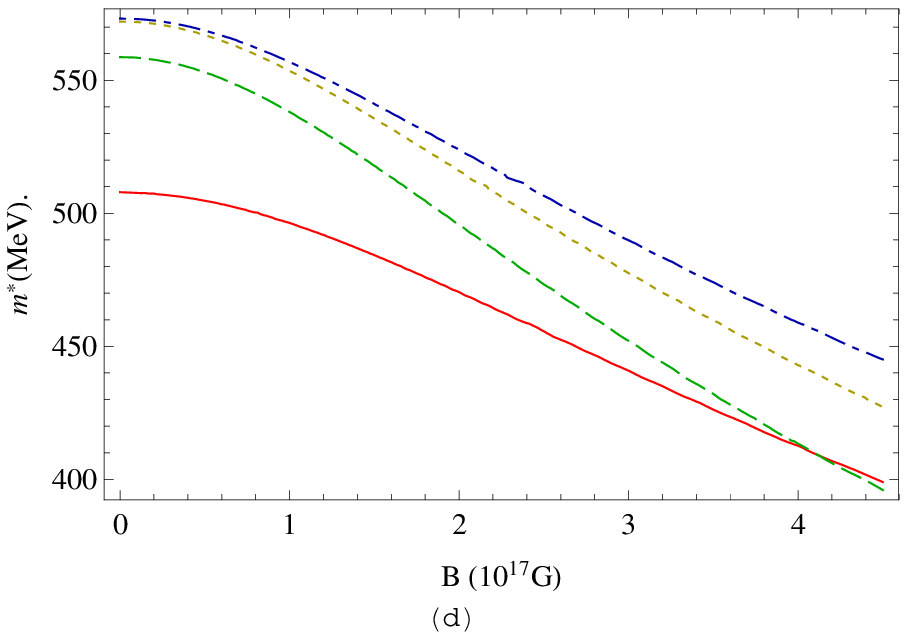}}\\
  \end{tabular}}
	\caption{ (Color online) Nuclear matter observables at saturation as a function of $B$: (a) the Fermi energy of the protons and neutrons (although only the neutron Fermi energy is shown since it does not noticeably differ from the proton values), while in (b) $n(\lambda)$ labels the Landau level occupation of the protons for each value of $\lambda$. Since $g_p=g_p^{(0)}<0$ the difference between $n(1)$ and $n(-1)$ is 1. Furthermore, since $B>0$ the $\lambda = 1$ protons have the lowest energies and thus occupy the most levels. In (c) and (d) the values of $g_v V_0$ and $m^*$ are plotted respectively. 
	}
	\label{fig:BEanal}
	\end{figure*}\\
	\\
While the various $E_b(B)$ curves are similar, the rate by which $\rho_0(B)$ increases differs. As shown in Fig.\ \ref{fig:BEanal}, the various baryon Fermi energies $\mu_b$ as well as the Landau occupation numbers $n(\lambda)$ do not differ significantly. Hence the $E_b(B)$ curves would also be similar since from Eq.\ (\ref{Eb}) we deduce that the binding energy goes like the Fermi energy. However, $m^*$ and $g_v V_0$ do vary significantly between the parameter sets. Since the underlying mechanisms are the same, these differences stem from the parametrization and must be the source of the variation in $\rho_0(B)$.\\
\\
For QHD1 $g_s\phi_0$ (from $m^*=m-g_s\phi_0$) and $g_v V_0$ have the largest values and so, since the Fermi energies are essentially the same, the QHD1 densities will be the lowest with the fewest number of occupied Landau levels, as we see from the $n(\lambda)$ plot. Applying the same logic to the rest of the plots, it is deduced that the FSU parameter set will have the highest values of $\rho_0(B)$.
\begin{figure*}
	\makebox[\textwidth][c]{%\centering	
  \begin{tabular}{ll}
    \includegraphics[width=.5\textwidth]{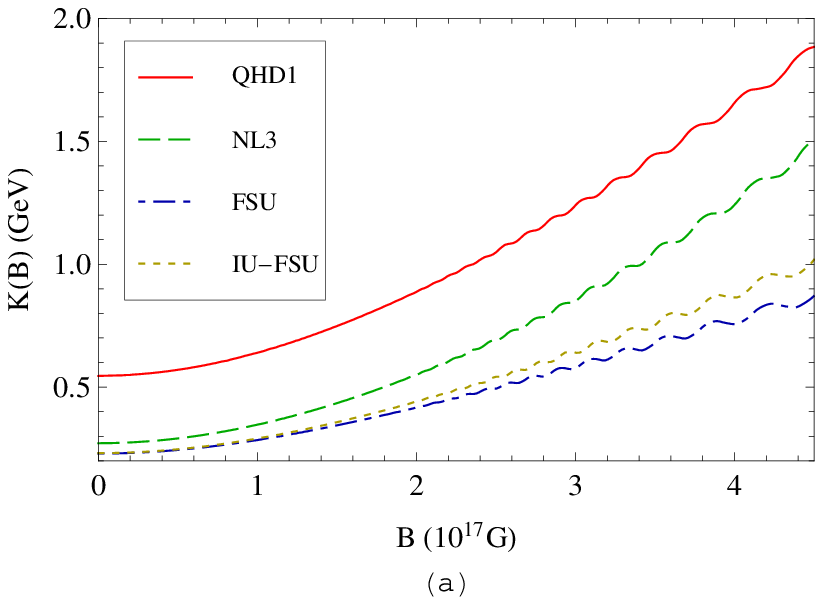}&
    \includegraphics[width=.5\textwidth]{{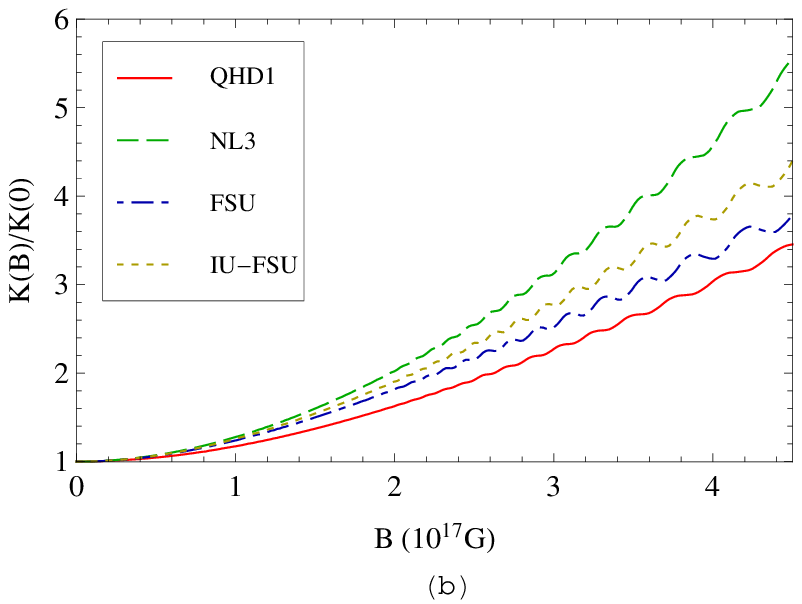}}
  \end{tabular}}
	\caption{ (Color online) The compression moduli of magnetized nuclear matter for $g_b=g_b^{(0)}$ for various QHD parameter sets are shown in (a), while in (b) the normalized values with regard to their values at $\rho_0(0)$ are shown.}
	\label{fig:K(B)}
	\end{figure*}
This is not surprising since in Ref.\ \cite{IUF} (and references therein) it has been well established that FSU has the softest equation of state (EoS). Since the EoS relates the energy density to the pressure a stiffer EoS exhibits a more rapid increase in pressure with density. Thus it can accommodate higher (energy) densities with the (comparatively) smallest increase in the pressure of the system.\\ 
\\
Related to the stiffness of the EoS is the compression modulus $K$ of nuclear matter which gives an indication of the compressibility of the matter. For magnetized matter the compression modulus $K(B)$ of the various QHD parameter sets was calculated using \cite{csg}
\begin{eqnarray}
	K(B) = 9\left[\rho^{2}\frac{d^{2}}{d\rho^{2}}\left(\frac{\epsilon}{\rho}\right)\right]_{\rho=\rho_{0}(B)}\label{K}.
\end{eqnarray}
We note that Eq.\ (\ref{K}) is essentially the derivative of the pressure as a function of density and thus an indication of the stiffness of the EoS. Since $\rho_0(B)$ increases with $B$, so too should $K(B)$ since the higher the density the more incompressible the system becomes. From Fig.\ \ref{fig:K(B)}(a) it is clear that QHD1 has the stiffest EoS while FSU has the softest. In Fig.\ \ref{fig:K(B)}(b) the normalized values of $K(B)$, with respect to $K(0)$, are shown.\\
\\
However, the increase in $K(B)$ does not happen smoothly, but rather it fluctuates with an increasing amplitude (for small $B$ these fluctuations are not visible on the scale of the figure). These fluctuations imply that as $B$ increases the system varies the degree to which it is compressible. This behavior is related to the number of Landau levels occupied by the system. 
\begin{figure*}
	\makebox[\textwidth][c]{%\centering	
  \begin{tabular}{ll}% Requires \usepackage{graphicx}
    \includegraphics[width=.5\textwidth]{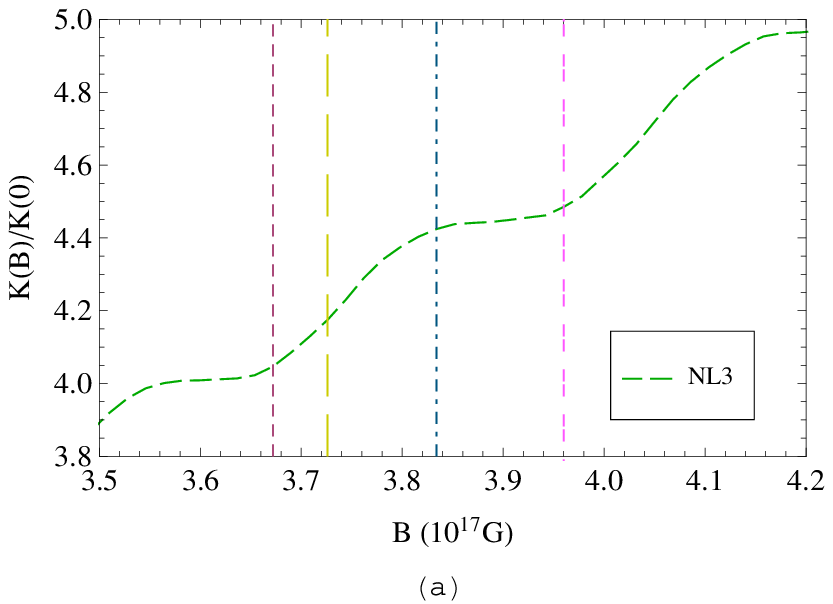}&
    \includegraphics[width=.5\textwidth]{{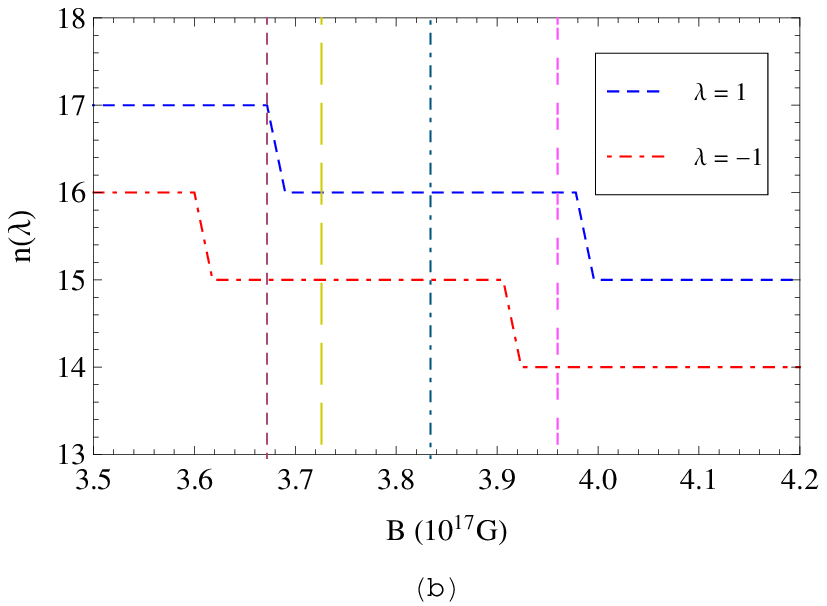}}
  \end{tabular}}
	\caption{ (Color online) Analysis of NL3 $K/K_0$ plot from Fig.\ \ref{fig:K(B)}(b) of which a section is shown in (a), while (b) shows the Landau level occupation. The vertical guidelines denote the same values of the magnetic field in both plots. }
	\label{fig:Kanal}
	\end{figure*}\\
	\\
In Fig.\ \ref{fig:Kanal} the relation between the fluctuations of the NL3 $K(B)$ and the depopulation \footnote{As $B$ increases the number of occupied Landau levels will decrease for a small change in density.} of the occupied Landau levels are shown. Variable behavior in a magnetized system of charged particles is not unexpected, since in a (two dimensional) quantum Hall system one observes dramatic variation of the conductivity due to the population of the Landau levels \cite{yobi}. Consequently the influence of the Landau levels on $K$ should not be entirely surprising. However, in this case the variability stems from the fact that the magnetic field influences the proton densities in two ways: through the degeneracy factor $\frac{|q_p B|}{4\pi^2}$ and the spacing between Landau levels which, from Eq.\ (\ref{Fp}), goes like $g_p B$.\\
\\
Therefore, as the magnetic field increases the degeneracy factor also increases which makes the system more compressible, but at the same time the energy gap between the Landau levels becomes bigger which makes the system more incompressible. Since the system is at its saturation density (hence in its lowest energy configuration), the Landau levels are depopulated one by one. As the occupation of the level with energy closest to the Fermi energy decreases the system becomes more compressible, since the particles from this level are absorbed by the Landau levels at lower energies.\\
\\ 
For neutrons $B\neq 0$ only induces a relative shift in the energy of particles with different orientations of their dipole moments. For $B>0$ the $\lambda=-1$ neutrons ($g_n>0$) become the dominant particle since the $\lambda=1$ neutrons will flip their dipole moments in order to attain a lower energy $\lambda=-1$ state. This contributes to the increase in $K(B)$, but has no influence on its fluctuations.\\ 
\\
To establish the asymmetric tendencies of a magnetized system the symmetry energy coefficient $a_4$ is calculated. It gives an indication of whether symmetric or asymmetric matter is preferred \cite{csg}. This coefficient is calculated at $\rho_0(B)$ from
\begin{eqnarray}
	a_{4} = \frac{1}{2}\left.\frac{\partial^{2}}{\partial t^{2}}\frac{\epsilon}{\rho_b}\right|_{t=0}\mbox{ with }\left(t\equiv\frac{\rho_{n} -
		 \rho_{p}}{\rho_b}\right)\label{a4}.
\end{eqnarray}
Since $a_4$ is the coefficient of the $\frac{(N-Z)^2}{N+Z}$ term in the semiempirical mass formula \cite{csg}, the larger $a_4$ becomes, the more symmetric matter is preferred in order to keep the energy at a minimum. \\
\\
The expression for $a_4$ can be simplified, but care should be taken since the baryon Fermi energies are also dependent on the magnetic field. A simplified expression for $a_4$ of magnetized matter is given in Ref.\ \cite{dienerPhD}. The results of the calculation are displayed in Fig.\ \ref{fig:a4(B)}.
\\
	\\
Similar to the other nuclear matter properties, $a_4$ increases with $B$ which indicates that more symmetric nuclear matter system is favored than when $B=0$. The increase in $a_4$ is intermittently interrupted by a slight decrease, signaling a preference for more asymmetric matter. Again this fluctuating nature of $a_4$ is related to the depopulation of Landau levels. Its origin is best illustrated by considering the density of the two types of protons ($\lambda=\pm1$), in particular the difference between the proton densities $\Delta\rho_p(\lambda)=\rho_p(1)-\rho_p(-1)$ which is shown in Fig.\ \ref{fig:a4(B)a}(b).\\
\\
For neutrons, $\Delta\rho_n(\lambda)$ increases linearly with $B$ since the difference in the energy of neutrons with different orientations of their dipole moment is $2g_n B$. For protons, $\Delta\rho_p(\lambda)$ is not directly proportional to $B$ since the difference in energy of protons with opposite signs of $\lambda$ depends on both $2g_p B$ as well as the number of filled Landau levels for a given $\lambda$. When $\Delta\rho_p(\lambda)$ is at a local minimum $\Delta n(\lambda)=1$ (which is the norm \footnote{In the Landau quantization the proton spectrum is such that the lowest energy proton level is not paired with a level of opposite $\lambda$ and hence $n(\lambda)$ differ by one (which choice of $\lambda$ has greater $n$ depends on the sign of $B$). See Ref.\ \cite{dienerPhD} for more details.}), while at a local maximum (just before a Landau level depopulates) $\Delta n(\lambda)=2$ and one choice of $\lambda$ proton levels is preferentially filled. In between these points in $\Delta\rho_p(\lambda)$,  $a_4$ decreases and more asymmetric matter is preferred since the degenerate proton Landau levels can be filled at a lower energy cost than the (nondegenerate) neutron energy levels.
\\
	\\
Of further interest is to investigate the properties of magnetized saturated nuclear matter for adjusted values of the baryon magnetic dipole moments, which are adjusted according to Eqs.\ (\ref{gbn}) and (\ref{gbp}). The strengths of the dipole moments were changed symmetrically so that the proton and neutron dipole moments are increased by the same factor; i.e., $g_b=10g_b^{(0)}$ means that the strength of both the dipole moments increased by a factor of 10. We observe that as $g_b$ is adjusted, the responses of the different QHD parameter sets are very similar. Hence only the results for the NL3 parameter set (an arbitrary choice) will be plotted.
\begin{figure*}%[tth]
	\makebox[\textwidth][c]{%\centering	
  \begin{tabular}{ll}
    	\includegraphics[width=.55\textwidth]{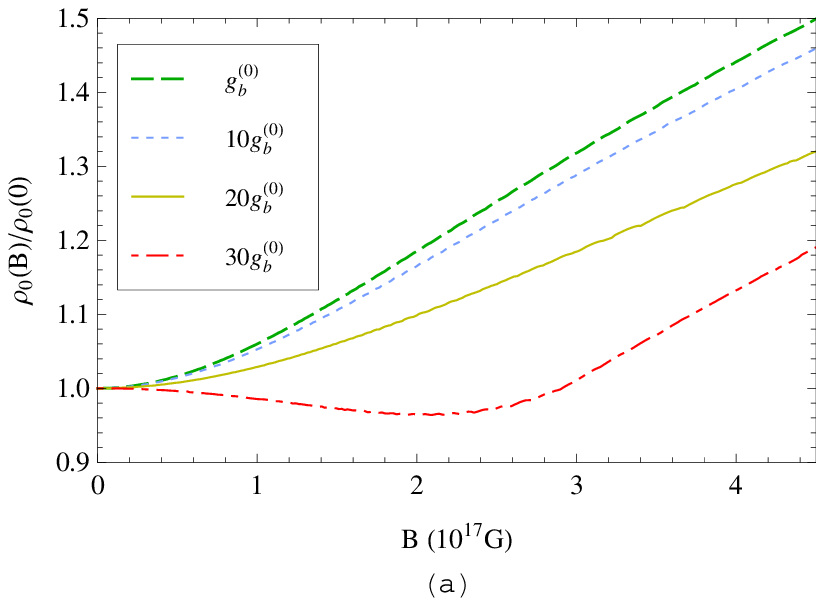}&
  	  \includegraphics[width=.5\textwidth]{{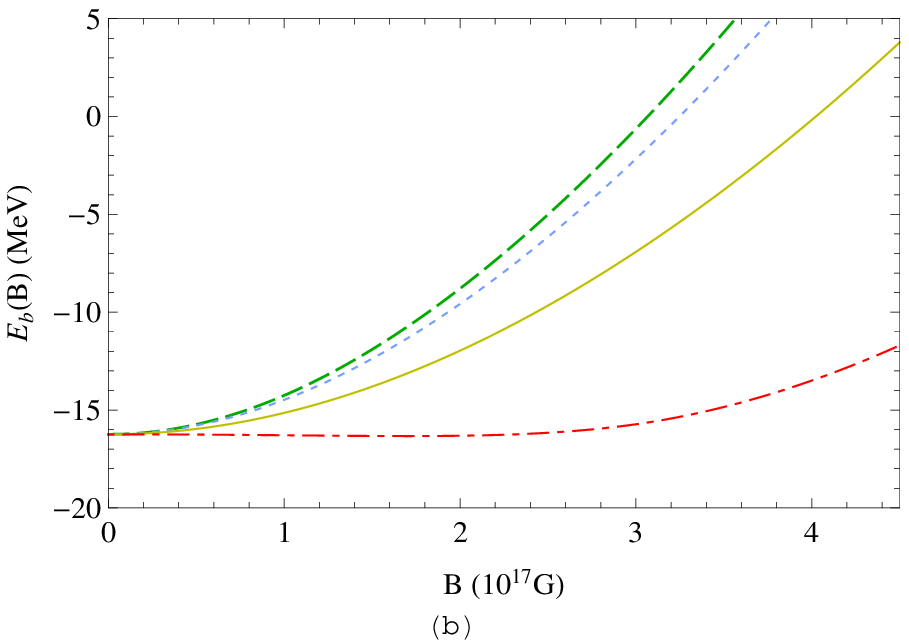}}\\
        	\includegraphics[width=.5\textwidth]{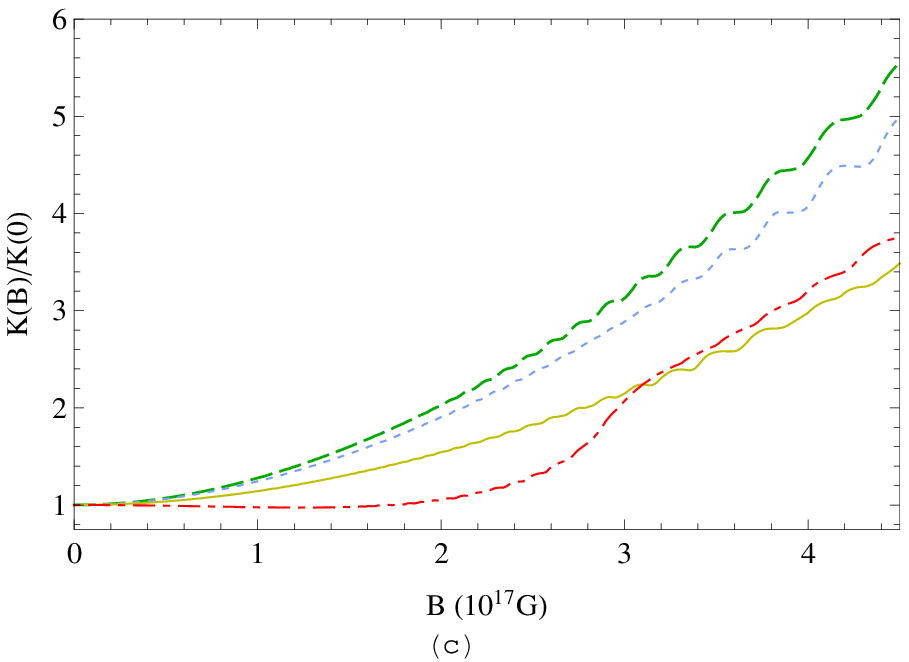}&
    	\includegraphics[width=.5\textwidth]{{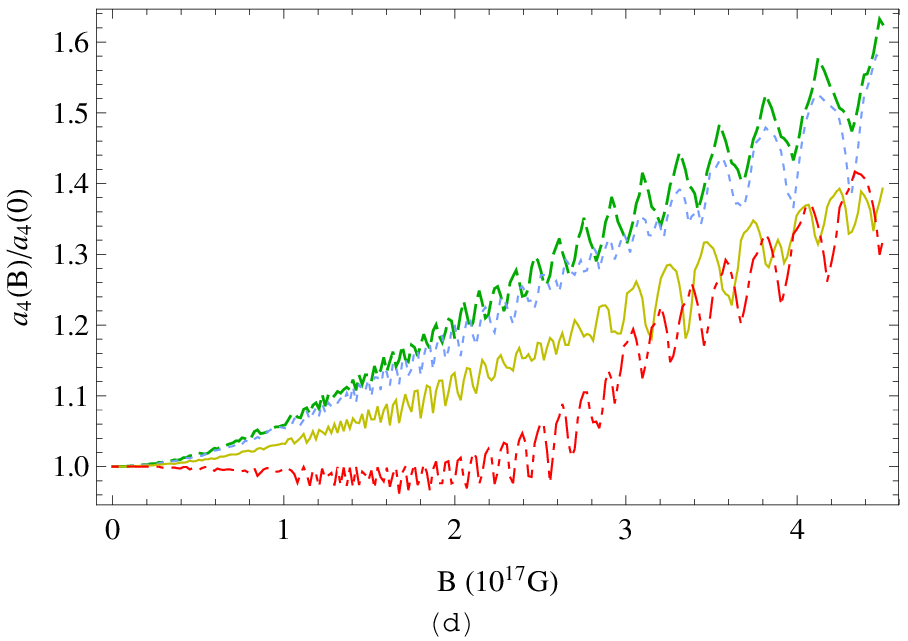}}
  \end{tabular}}
	\caption{ (Color online) Magnetized nuclear matter properties for various values of $g_b$, as multiples of $g_b^{(0)}$: (a) the saturation densities, (b) the binding energies, (c) $K(B)$, and (d) $a_4(B)$.	}
	\label{fig:BEgn}
	\end{figure*}\\
\\
In Fig.\ \ref{fig:BEgn} the values of the different properties for $g_b$ equal to $g_b^{(0)}$, $10g_b^{(0)}$, $20g_b^{(0)}$, and $30g_b^{(0)}$ are shown. We do not claim that these values of $g_b$ are necessarily feasible or attainable for the plotted range of densities and magnetic fields, but rather that they illustrate the full spectrum of the possible behavior of magnetized matter under extreme conditions.\\
\\
It is observed that as $g_b$ increases the system becomes less dense but more bound. The density decreases since the gap between different dipole orientations of the protons and the neutrons, as well as the separation of the proton Landau levels, increases. Hence lower energies are attained at the cost of the number of particles that the system can accommodate per unit volume. For very large values of $g_b$ the relative shift between the different $\lambda$ proton and neutron energy levels is so large that only one choice of $\lambda$ (the one with the lowest energy levels) is populated. \\
	\\
The bottom row in Fig.\ \ref{fig:BEgn} shows $a_4(B)$ and $K(B)$ as $g_b$ varies for the NL3 parameter set. For both the fluctuating behavior persists but their respective increases become less rapid as $g_b$ increases, which mimics the manner in which $\rho_0(B)$ changes with $g_b$. For $g_b=30g_b^{(0)}$ the last $\lambda=-1$ proton Landau level depopulates at $B\approx3\times 10^{17}$ G . Leading to this point both $K$ and $a_4$ increase: $K$ increases since the system becomes much more incompressible because almost only $\lambda=1$ Landau levels are filled. On the other hand $a_4$ shifts from asymmetric (proton-rich) matter to more symmetric matter since only $\lambda=1$ protons are accommodated.
\section{Discussion}
\begin{figure*}
	\makebox[\textwidth][c]{	
  \begin{tabular}{ll}
    \includegraphics[width=.5\textwidth]{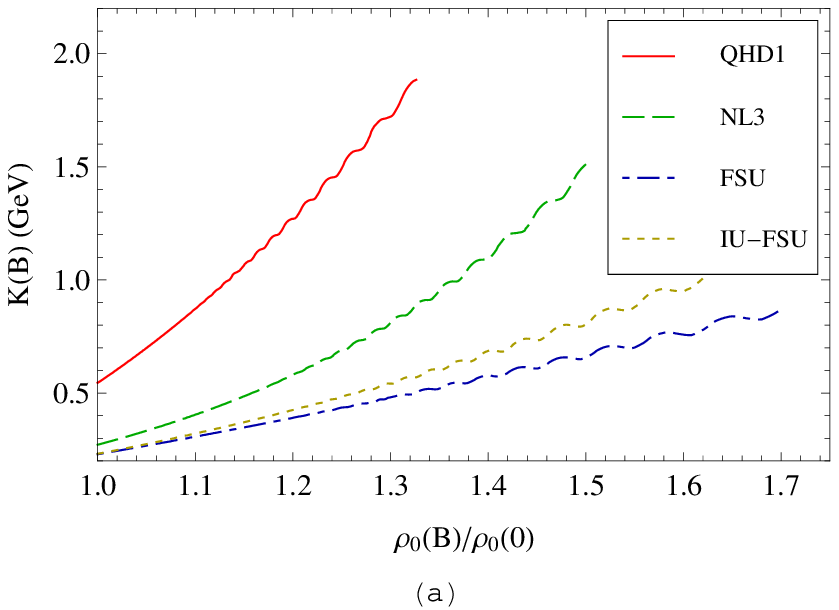}&
    \includegraphics[width=.5\textwidth]{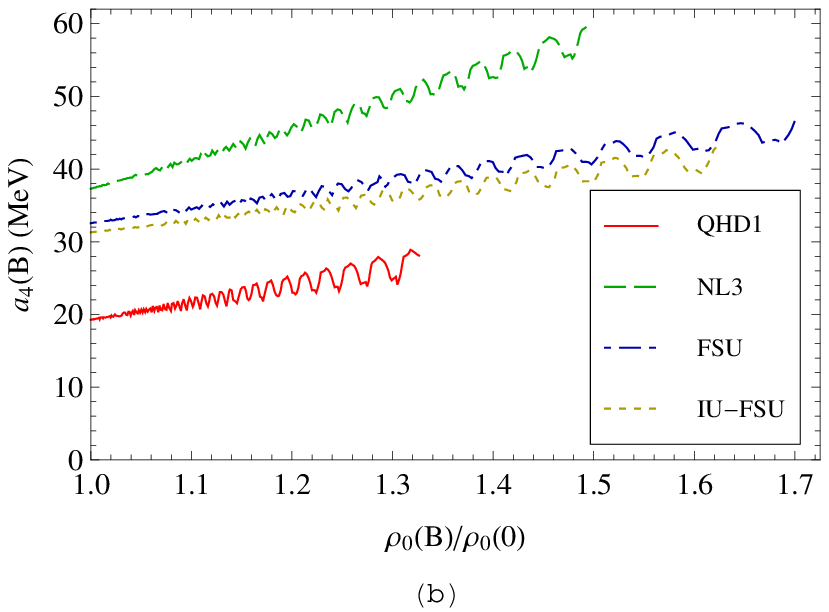}
  \end{tabular}
	}
	\caption{ (Color online) (a) $K(B)$, and (b) $a_4(B)$ as a function of $\rho_0(B)$ for $g_b=g_b^{(0)}$.}
	\label{fig:satrho}
\end{figure*}

From our results we conclude that for $B< 10^{15}$ G the saturation properties of symmetric nuclear matter are not substantially influenced. However, as the value of $q_b B$ and/or $g_b B$ comes (relatively) close to that of the baryon (reduced) mass the system is undeniably influenced by the magnetic field: For $g_b=g_b^{(0)}$ and $B=10^{15}$ G, $g_bB\approx 3$ keV and $q_p B\approx 30$ keV while $m^*$ ranges between 500 and 580 MeV, depending on the parameter set. Consequently we find that, similar to nuclei \cite{penaA}, magnetic fields of the order of $10^{16}$ G and up are needed to influence the properties of saturated symmetric nuclear matter. However, it is important to note that this is a question of energy scales (in particular the relation of $m^*$ to $q_pB$), rather than absolute value of the magnetic field strength.\\
\\
To establish the scope of the magnetic field's influence, it is useful to plot $K(B)$ and $a_4(B)$ as a function of $\rho_0(B)$, shown in Fig.\ \ref{fig:satrho}.
Shown there are the properties of saturated nuclear matter for an increasing magnetic field. As the magnetic field increases, so too does the saturation density. Although the saturation density is still at the minimum of the binding energy curve, the matter at this density becomes less bound as the magnetic field and correspondingly $\rho_0(B)$ increase, as was shown in Fig.\ \ref{fig:BE(B)}(b).\\
\\
The overall tendency of saturated matter to become more incompressible as the density increases is expected, but the smaller scale fluctuations in the increase are surprising. These correlate with the depopulation of the proton Landau levels. When the Landau levels depopulate the matter becomes more compressible despite the increase in the density.\\
\\
As $B$ and $\rho_0(B)$ increase, $a_4$ also shows an overall increasing trend. As mentioned previously, the larger $a_4$ becomes the more symmetric matter is preferred, according to the semiemperical mass formula \cite{csg}. From this we deduce that for higher $B$ nuclear matter tends to be more symmetric, thus having a larger proton fraction than expected since the degenerate Landau levels can accommodate particles at a lower energy cost. %for unmagnetized matter. 
Once again the increase in $a_4$ is not smooth and there are sections where, despite the increase in $\rho_0(B)$, more asymmetric (neutron rich) matter is preferred. These are also shown to be related to the depopulation of proton Landau levels.\\
\\
Hence it is obvious that any sudden change in the magnetic field will drastically alter the properties of the saturated nuclear matter, since the configuration of the Landau levels will change. These changes have a direct impact on the compressibility of the system as well as the preferred mix of protons and neutrons. In Fig.\ \ref{fig:BEgn} we also showed that the influence of the magnetic field persists when the magnetic dipole moment increases. \\
\\
In a naive way Fig.\ \ref{fig:satrho} can be seen as indicative of the behavior of nuclear matter in a section of the magnetar interior, assuming that the magnetic field increases with the depth. We deduce that the compressibility as well as proton fraction of nuclear matter in the stellar interior is dependent on the magnetic field. Furthermore, if the magnetic field were to suddenly change, it could potentially significantly alter its nuclear properties.\\
\\
Highly magnetized neutron stars (magnetars) are known for various types of flares/bursting activities (see Ref.\ \cite{chap14} for a review), some of which that are accompanied by glitches (sudden increase in the rotation frequency of the star) \cite{AXPprogress}. It has been reported by Woods {\em et al.} in Ref.\ \cite{Brecon} that the flare in magnetar SGR 1900+14 was accompanied by the reconfiguration of the stellar magnetic field. Hence the changes in the stellar magnetic field, as well as the accompanying changes in the conditions within neutron star interior, may contribute to the observed properties of magnetars. \\
\\
If the reconfiguration of the magnetic field implies a change in the magnetic field strength throughout the interior, the changes in the compressibility can induce a compression wave since the compressibility depends on the magnetic field in a nonlinear, fluctuating fashion. If the changes in the magnetic field happen over a short enough timescale, differential stresses will build up in the interior due to the varying nature of $K$.\\
\\
From $a_4(B)$ in Fig.\ \ref{fig:satrho}(b) changes in $B$ might imply a change in the preferred composition of nuclear matter with regards to the ratio of protons and neutrons. Hence a change in $B$ could induce inverse or normal beta decay in nuclear matter. Not only would such decay activities change the composition of nuclear matter, but would also release energy which could be the source of some of the sudden bursts of radiation observed from magnetars. 
\section{Conclusion}
We have presented some evidence that the composition and properties of magnetized, saturated symmetric nuclear matter start to depend on $B$ for $B \gtrsim 10^{16}$ G. We have also shown that the influence of the magnetic field persists even if the strength of the baryon magnetic dipole moments changes.\\
\\
We believe that the influence of the magnetic field is important when studying the magnetar interior. However, since the internal dynamics of the neutron star interior and, in particular, the time dependence of the nuclear processes are not known it is difficult to model dynamical behavior. If the origin and behavior of the stellar magnetic field is known, then we might have a chance of calculating the influence of any changes on the interior. In a future publication we will consider whether a ferromagnetic phase might be present in the neutron star interior as the source of the magnetar magnetic field. 
\section{Acknowledgements}
This research is supported by the South African SKA project as well as the National Research Foundation of South Africa.
\bibliography{fskryfnuc}
\end{document}